# Enhancement of the critical current density and flux pinning of $MgB_2$ superconductor by nanoparticle SiC doping

S.X. Dou[1]*, S. Soltanian[1], J. Horvat[1], X.L. Wang[1], P. Munroe[2], S.H. Zhou[1], M. Ionescu[1], H. K. Liu[1] and M. Tomsic[3]

[1]*Institute for Superconducting and Electronic Materials, University of Wollongong, Northfields Ave. Wollongong, NSW 2522, Australia*
[2]*Electron Microscopy Unit, University of New South Wales, Sydney, NSW 2001, Australia*
[3]*Hyper Tec Research Inc., 110 E. Canal St., Troy, OH 45373, USA*

Abstract
Doping of $MgB_2$ by nano-SiC and its potential for improvement of flux pinning was studied for $MgB_{2-x}(SiC)_{x/2}$ with x = 0, 0.2 and 0.3 and a 10wt% nano-SiC doped $MgB_2$ samples. Co-substitution of B by Si and C counterbalanced the effects of single-element doping, decreasing $T_c$ by only 1.5K, introducing pinning centres effective at high fields and temperatures and enhancing $J_c$ and $H_{irr}$ significantly. Compared to the non-doped sample, $J_c$ for the 10wt% doped sample increased by a factor of 32 at 5K and 8T, 42 at 20K and 5T, and 14 at 30K and 2T. At 20K, which is considered to be a benchmark operating temperature for $MgB_2$, the best $J_c$ for the doped sample was $2.4 \times 10^5 A/cm^2$ at 2T, which is comparable to $J_c$ of the best Ag/Bi-2223 tapes. At 20K and 4T, $J_c$ was $36,000 A/cm^2$, which was twice as high as for the best $MgB_2$ thin films and an order of magnitude higher than for the best $Fe/MgB_2$ tapes. Because of such high performance, it is anticipated that the future $MgB_2$ conductors will be made using the formula of $MgB_xSi_yC_z$ instead of the pure $MgB_2$.

* Correspondence should be addressed to S. X. Dou( e-mail: shi_dou@uow.edu.au)

The critical current density ($J_c$) in $MgB_2$ has been a central topic of research since superconductivity in this compound was discovered [1]. High $J_c$ values of $10^5$ to $10^6$ $A/cm^2$ have been reported for $MgB_2$ by several groups [2-11]. However, $J_c$ drops rapidly with increasing magnetic field due to its poor flux pinning. To take advantage of its high $T_c$ (39K), improvement of irreversibility field ($H_{irr}$) and $J_c(H)$ was achieved by oxygen alloying of $MgB_2$ thin films [12] and by proton irradiation of $MgB_2$ powder [13]. However, for practical applications, the pinning centres should be introduced by a simple process, like chemical doping.

Most of element substitution studies were aimed at rising $T_c$, which were limited to a low doping level. High level doping resulted in multiple phases. Improvement of flux pinning was also attempted, using doping by chemical compounds. Here, the results are largely limited to the addition of the compounds, rather than substitution of Mg or B by the compounds. The additives alone appear to be ineffective for improvement of pinning at high temperatures. Zhao et al., and Feng et al. doped $MgB_2$ with Ti and Zr, improving $J_c$ at 4K, attributable to the sintering aid effect by these additives [14,15]. At temperatures above 10K, $J_c$ dropped off rapidly with increasing field ($H_{irr}$ = 4T at 20K). Recently, Wang et al. doped $MgB_2$ with nano-particle $Y_2O_3$ [16]. They obtained a significant improvement of irreversibility field ($H_{irr}$= 11.5T) at 4.2K, due to the introduction of dispersed inclusions, such as $YB_4$. However, the improvement was insignificant at 20K ($H_{irr}$ = 5.7T). Cimberle et al., found that doping with a small amount of Li, Al and Si showed some increase in $J_c$, but there was no improvement in $H_{irr}$ [17], because single element doping degraded $T_c$ dramatically at high doping level. Slight improvements in $J_c(H)$ were achieved by various mechanical deformation processes in hot-pressed bulk, wires and tapes [4-12, 18]. Because the $MgB_2$ lattice structure is rather rigid and it contains only two elements, the density of thus introduced defects was too low for effective pinning.

Recently, we found that chemical doping of nano-SiC into $MgB_2$ can significantly enhance $J_c$ in high fields with only slight reduction of $T_c$ up to the doping level of 40% of B [19]. This finding clearly demonstrated that co-substitution of SiC for B in $MgB_2$ induced intra-grain defects and high density of nano-inclusions as effective pinning centres. However, the processing conditions used were far from optimised and the sample density was only 50% of the theoretical value. Therefore, it was not possible to assess the full potential of nano-SiC doping in improvement of $J_c$. In this work, we show that the nanometer size SiC doped $MgB_2$ gives the highest $J_c$ values in high magnetic fields at 20K ever reported for any form of $MgB_2$, including the thin films.

$MgB_2$ pellet samples were prepared by an in-situ reaction method, which was described in details previously[20]. Powders of magnesium (99% pure) and amorphous boron (99% pure) were well mixed with SiC nano-particle powder (size of 10nm to 100nm) with the atomic ratio of $MgB_{2-x}(SiC)_{x/2}$, where x = 0, 0.2, and 0.3, for samples designated as 1 to 3, respectively. A sample with 10 wt% of SiC addition to $MgB_2$ was also made as sample 4. The pellets 10 mm in diameter and 2 mm in thickness were made under uniaxial pressure, sealed in the Fe tube and then heated at temperatures 700-900°C for 1 hour in flowing high purity Ar. This was followed by furnace cooling to room temperature. The magnetization of samples was measured over a temperature range of 5 to 30 K using a PPMS (Quantum Design) in a time-varying magnetic field with sweep rate 50 Oe/s and amplitude up to 8.5T. All the samples were cut to the same size of 0.56x2.17x3.73 mm$^3$ from as-sintered pellets. A magnetic $J_c$ was derived from the height of the magnetization loop $\Delta M$ using a Bean model: $J_c=20\Delta M/[a(1-a/3b)]$. $J_c$ versus magnetic field up to 8.5 T was measured at 5, 10, 15, 20, and 30 K. Irreversibility field ($H_{irr}$) was obtained from measuring the field-cooled (FC) and zero-field-cooled (ZFC) magnetic moment as a function of temperature for several values of the field. The temperature at which the FC and ZFC branches met within the experimental uncertainty (corresponding to $J_c$ of 100 A/cm$^2$) was the irreversibility temperature for the particular field used. Critical temperature ($T_c$) was obtained as the onset of diamagnetic transition in magnetic ac susceptibility measurements.

Fig. 1 shows XRD patterns for the SiC doped and non-doped samples. The XRD pattern for the non-doped sample (sample 1) reveals about 5% MgO, beside $MgB_2$ as the main phase. Samples 2 and 3 consist of $MgB_2$ as the main phase, with $Mg_2Si$ as the major impurity phase (crosses in Fig. 1). The estimated fraction of $Mg_2Si$ was 10%. The EDS analysis results showed that Mg:Si ratio was identical over all the sample area, indicating a homogeneous phase distribution.

Value of $T_c$ for the non-doped sample was 37.6K. For the doped samples, $T_c$ decreased with increasing doping level. The transition width was typically 0.5K. It is striking to note that $T_c$ only dropped by 1.3K for doping level of x=0.3. This corresponds to 15wt.% of SiC in $MgB_2$. In contrast, $T_c$ was depressed by almost 7K for 10wt% C substitution for B in $MgB_2$ [21]. According to Cimberle et al., 0.5at% Si substitution reduced $T_c$ by about 0.5K [17]. These results suggest that the higher tolerance of $T_c$ of $MgB_2$ to SiC doping is attributable to the co-substitution of B by C and Si. This is because the atomic radius of C (0.077nm) and Si (0.11nm) atoms is close to that of B (0.097nm). Co-doping with SiC counterbalanced the negative effect on $T_c$ of the single element doping.

Fig. 2 shows $J_c(H)$ curves for the SiC-doped $MgB_2$ samples at 5K, 20K, and 30K, for different doping levels. It is noted that all the $J_c(H)$ curves for doped samples show a crossover with the non-doped samples at higher fields. Although SiC doping caused a slight reduction of $J_c$ in low fields, it is much larger than for the non-doped ones in high fields for all the measured temperatures. Compared to the non-doped sample, $J_c$ for the 10wt% doped sample increased by a factor of 32 at 5K and 8T, 23 at 15K and 6T, 42 at 20K and 5T, and 14 at 30K and 2T. This is the best $J_c(H)$ performance ever reported for $MgB_2$ in any form. It is noted that the $J_c(H)$ curves for the non-doped sample showed a rapid drop in high fields and a plateau near $H_{irr}$. Earlier, we ascribed this phenomenon to the grain decoupling at higher fields, as a consequence of impurities at the grain boundaries [4]. In contrast, all the SiC doped samples do not show this phenomenon, indicating the grain decoupling occurs at higher fields, as either the substitution or the induced nano-inclusions are incorporated into the grains.

Fig. 3 shows a comparison of $J_c(H)$ for 10 wt% SiC doped sample at 20K with data reported in literature. $J_c$ for this sample exhibits better field performance and higher values of $J_c$ in high field than any other element doped samples [14-16] and the non-doped wires [22]. Our SiC doped $MgB_2$ are even better than the thin film $MgB_2$ (Fig. 3), which exhibited the strongest reported flux pinning and the highest $J_c$ in high fields to date. At 20K, the best $J_c$ for the 10wt% SiC doped sample was $10^5 A/cm^2$ at 3T, which exceeded $J_c$ of the state-of-the-art Ag/Bi-2223 tapes. At 20K and 4T, $J_c$ was $36,000 A/cm^2$, which is twice as high as for the best $MgB_2$ thin films [12] and an order of magnitude higher than for the state-of-the-art $Fe/MgB_2$ tapes [10].

Temperature dependence of $H_{irr}$ for nano-SiC doped $MgB_2$, as well as for the pellets and tapes prepared previously (round symbols), is shown in the inset to Fig 3. Apparently, $H_{irr}$ for x=0 overlaps with $H_{irr}$ for the previous samples, even though the latter had significantly smaller values of $J_c$. Doping with SiC significantly improved $H_{irr}$. For example, $H_{irr}$ for SiC doped sample reached 7.3T at 20K, compared to 5.7T for the non-doped one. This is consistent with improvement of field dependence of $J_c$ with the doping. Because $H_{irr}$ for the non-doped control sample (x=0) is the same as for the previously prepared samples, the improvement of $J_c(H)$ occurred indeed because of the improvement of flux pinning by the doping and not because of improved sintering of $MgB_2$. Regarding the mechanism of the enhancement of $J_c$ at higher fields, it is necessary to recognize the special features of SiC doping. First, in contrast to previous work on doping for improving $J_c$ [14-16, 18], SiC doping has no densification effect, as evidenced by the fact that the density of doped samples is $1.2 g/cm^3$, independent of the doping level. In addition, SiC doping takes place in the form of substitution and/or addition [19], while in the previously reported work [14-16, 18], the doping was in the form of additives, not incorporated into crystalline lattice of $MgB_2$.

The TEM images showed a high density of dislocations and a large number of ~10nm inclusions inside the grains (Fig. 4). Their concentration increased with the doping level. EDS analysis of the grains revealed a presence of uniformly distributed Mg, B, C, Si and O (inset to Fig. 4). This, and the results of XRD, suggests that the inclusion nano-particles were made of $Mg_2Si$, or unreacted SiC. All the intra-grain defects and the inclusions act as effective pinning centres. Our results suggest that a combination of substitution-induced defects and highly dispersed additives are responsible for the enhanced flux pinning. When SiC reacts with liquid Mg and amorphous B at the sintering temperatures, the nano-particles of SiC will act as nucleation sites to form $MgB_2$ and other non-superconducting phases and some nano-particles can be included within the grains as inclusions. Thus, the reaction-induced products are highly dispersed in the bulk matrix.

Given the ease of production of SiC-doped $MgB_2$, our results significantly strengthen the position of $MgB_2$ as a competitor to more expensive conventional superconductors and HTS. It is evident that the future $MgB_2$ conductors will be made using a formula of $MgB_xSi_yC_z$ instead of pure $MgB_2$, because SiC doping is easily achievable and results is strong improvement of flux pinning. In the present study, the density of the pellet samples is still very low, only about $1.3 g/cm^3$. Thus, still higher $J_c$ can be achieved by improving the density.

In summary, we have demonstrated that the critical current density, irreversibility field and flux pinning of nano-SiC doped $MgB_2$ in bulk form is higher than any other type of $MgB_2$ reported so far, paving the way for $MgB_2$ to potentially replace the current market leader in superconducting applications, Nb-Ti.

Acknowledgment

The authors thank Drs. T. Silver, M.J. Qin, A. Pan, E.W. Collings, M. Sumption and Mr R. Neale for their helpful discussion. This work was supported by the Australian Research Council, Hyper Tech Research Inc., OH, USA, Alphatech International Ltd, NZ and the University of Wollongong

Fig. 1: X-ray diffraction patterns for the non-doped and SiC-doped samples

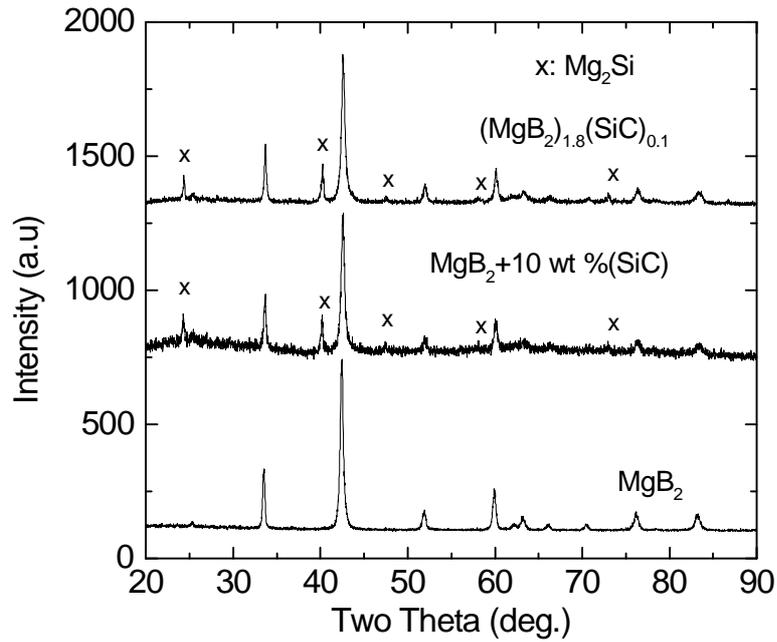

Fig. 2: The $J_c(H)$ dependence at 5, 20 and 30 K for samples 1, 2, 3, and 4 shown by solid, dashed and dotted line, and crosses, respectively.

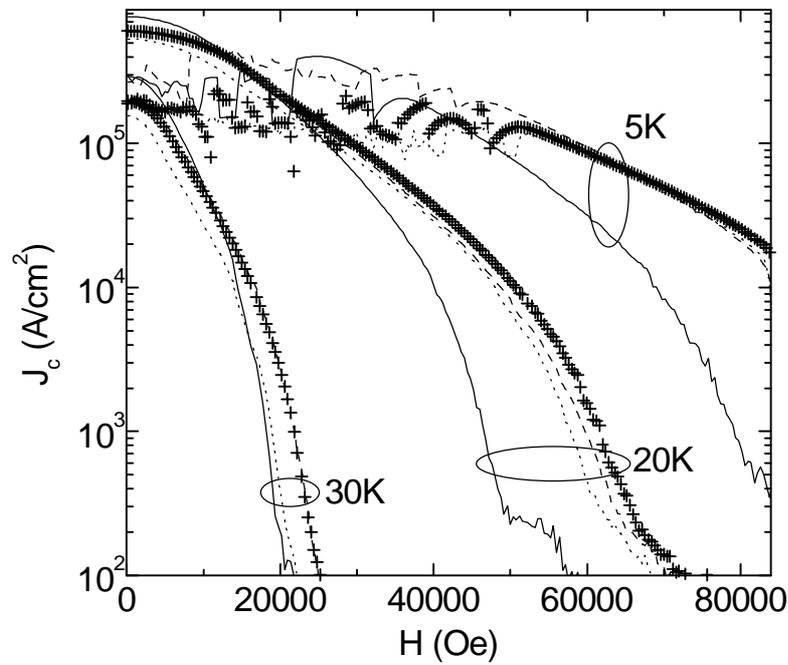

Fig. 3. A comparison of $J_c(H)$ at 20K for the 10wt% SiC doped sample (sample 4) and for Ti doped [14], $Y_2O_3$ doped [16], thin film with strong pinning [12] and Fe/$MgB_2$ tape [22], which represent the state-of-the-art performance of $MgB_2$ in various forms. Inset: temperature dependence of the irreversibility field for SiC doped $MgB_2$ with different content of SiC (triangles and squares) and for previously prepared non-doped $MgB_2$ (round symbols).

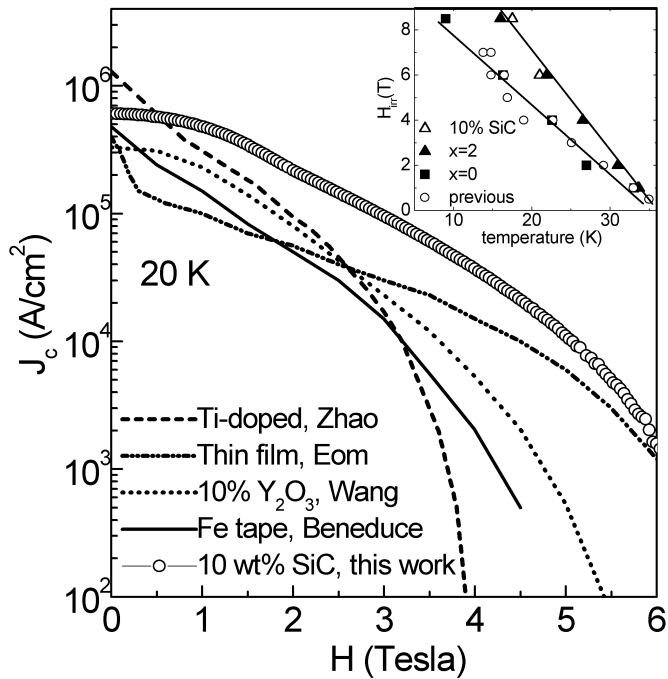

Fig. 4: TEM image showing the intra-grain dislocations and nano-particle inclusions within $MgB_2$ grains. Inset: the EDS element analysis of $MgB_2$ grains.

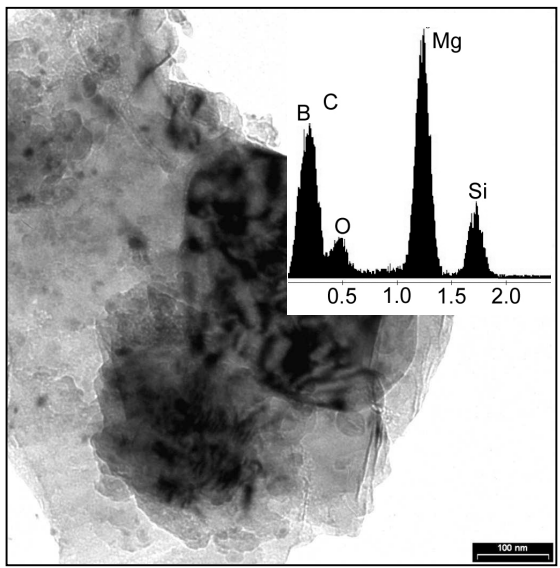